\documentclass[acmsmall,screen,nonacm]{acmart}

\usepackage{amsmath}
\usepackage{mathtools}
\usepackage{booktabs}
\usepackage{tabularx}
\usepackage{listings}
\usepackage{placeins}
\newsavebox{\openlanguagebox}
\setcopyright{none}
\authorsaddresses{}
\renewcommand\footnotetextcopyrightpermission[1]{}
\newtheorem{theorem}{Theorem}
\newcommand{\rir}{\textsc{RIR}}
\newcommand{\ov}{\textsc{OracqVerify}}
\newcommand{\den}[1]{\llbracket #1\rrbracket}
\newcommand{\ket}[1]{\lvert #1\rangle}
\newcommand{\opnorm}[1]{\left\lVert #1\right\rVert_{\mathrm{op}}}
\newcommand{\ctrl}{\operatorname{ctrl}}
\newcommand{\Inline}{\operatorname{inline}}

\title{Verifying Quantum Programs Before Oracle Binding}
\author{Zhao-Yun Chen}
\authornote{Correspondence to Zhao-Yun Chen (\nolinkurl{chenzhaoyun@iai.ustc.edu.cn}) and Yu Zhang (\nolinkurl{yuzhang@ustc.edu.cn}).}
\email{chenzhaoyun@iai.ustc.edu.cn}
\affiliation{%
  \institution{Institute of Artificial Intelligence, Hefei Comprehensive National Science Center}
  \city{Hefei}
  \state{Anhui}
  \country{China}
}

\author{Ming-Yang Tan}
\affiliation{%
  \institution{Laboratory of Quantum Information, University of Science and Technology of China}
  \city{Hefei}
  \state{Anhui}
  \country{China}
}

\author{Hao Pan}
\affiliation{%
  \institution{School of Computer Science and Technology, University of Science and Technology of China}
  \city{Hefei}
  \state{Anhui}
  \country{China}
}

\author{Yu Zhang}
\authornotemark[1]
\email{yuzhang@ustc.edu.cn}
\affiliation{%
  \institution{School of Computer Science and Technology, University of Science and Technology of China}
  \city{Hefei}
  \state{Anhui}
  \country{China}
}

\author{Guo-Ping Guo}
\affiliation{%
  \institution{Laboratory of Quantum Information, University of Science and Technology of China}
  \city{Hefei}
  \state{Anhui}
  \country{China}
}
\affiliation{%
  \institution{Anhui Province Key Laboratory of Quantum Network, University of Science and Technology of China}
  \city{Hefei}
  \state{Anhui}
  \country{China}
}
\affiliation{%
  \institution{Institute of Artificial Intelligence, Hefei Comprehensive National Science Center}
  \city{Hefei}
  \state{Anhui}
  \country{China}
}
\affiliation{%
  \institution{Origin Quantum Computing Technology (Hefei) Co., Ltd.}
  \city{Hefei}
  \state{Anhui}
  \country{China}
}

\begin{document}
\begin{abstract}
Oracle-based quantum algorithms specify how the circuit can access its input through coherent operations. The correctness of the algorithm can be established in terms of these operations before their implementations are chosen. This paper presents a register-level format for such open quantum programs and a method for verifying their late bindings. The format represents unresolved oracle calls, ordered register views, read-only quantum random-access memory (QRAM), and private workspace. Our validator compares independently supplied open and bound programs. We prove in Lean that an accepted binding transfers the caller's proof when the chosen implementations meet their public matrix contracts and restore private workspace to zero. The theorem accounts for coherent control and adjoints. Quantitative variants bound error for approximate oracle implementations and for callers with a certified projected denotation. OracqVerify implements the checker for serialized Oracq programs. We demonstrate verification on gate and QRAM bindings for Grover search, Boolean oracles for Deutsch--Jozsa, and alternative Hamiltonian simulation components for a fixed evolution caller.
\end{abstract}

\maketitle

\section{Introduction}

Quantum algorithm design routinely separates the algorithm from the operations through which it accesses its input~\cite{aaronson2021open}. In an \emph{oracle-based} formulation, an instance is supplied through one or more \emph{oracles}, operations with a specified action that the algorithm can invoke coherently on superpositions. The circuit interleaves oracle calls with gates specified independently of the instance, and its correctness is proved for instances satisfying the problem's promises. Query complexity counts the oracle calls. This gives a common model for algorithmic upper bounds and lower bounds, independently of the cost of realizing each oracle on a particular machine~\cite{beals2001quantum}.

The access model also determines which algorithmic constructions can be reused. Sparse Hamiltonian simulation and quantum linear-system algorithms formulate their matrix access in terms of coherent queries for entry locations and values. The same construction can then be analyzed for a family of matrices equipped with those operations~\cite{berry2015hamiltonian,childs2017linear}. An operator can also be supplied through a \emph{block encoding}, a unitary with a designated submatrix equal to the operator up to a known scale. Quantum singular value transformation uses calls to such a unitary and its adjoint to encode admissible polynomial transformations of the operator's singular values. Its analysis depends on the block-encoding specification, which allows the same transformation to be used with different circuits for supplying the operator~\cite{gilyen2019quantum}.

To turn this separation into a software architecture, a program must be able to retain its oracle calls while deferring their implementation. A matrix-entry oracle, for instance, can compute an entry reversibly from a formula or coherently read it from a table. These choices can supply the same logical query with different gate, workspace, and memory costs. The choice can be made once the input representation and available memory are known. Oracq supports this workflow by allowing a caller to be generated and serialized with named oracle slots. Supplying implementations for those slots after constructing the caller is \emph{late binding}~\cite{chen2026oracq}.

At the binding boundary, the assumptions of the algorithmic proof become obligations on the executable program. Each call must act on the intended ordered register views and use the intended memory bank, and each implementation must realize its specified public action and return private workspace to zero. The caller may also invoke the operation under coherent control or take its adjoint. For example, \(U\) and \(-U\) describe the same channel in isolation and induce different relative phases under coherent control.

We study how to verify this relationship between an open program and an independently supplied bound program. The objective is to transfer a proof about the open caller to the executable result, using separate evidence for the structure of the binding and for the semantics of the chosen implementations. We make two main contributions.

\begin{enumerate}
\item \textbf{A format for open and bound quantum programs.} We specify and formalize the fragment of Oracq's register-level intermediate representation (\rir{}) used for late binding. The format records oracle slots, ordered register views, scoped workspace, and read-only memory arguments. It represents the caller before an implementation is chosen and records how each call reaches its implementation after binding.
\item \textbf{Verified late binding.} An executable validator checks caller structure, wire order, and bank forwarding in the decoded open and bound programs. A Lean linking theorem transfers the open proof across an accepted binding when the implementations satisfy their public matrix contracts and have clean exits. Quantitative rules bound accumulated operator-norm error for approximate public actions and give weighted bounds for callers with certified projected denotations.
\end{enumerate}

We implement these contributions in \ov{}. Grover search demonstrates checking alternative gate and QRAM bindings. Deutsch--Jozsa demonstrates certified generation and independent verification at arbitrary positive address widths. In linear combination of Hamiltonian simulation (LCHS), we check three bindings of the same open caller, substituting Taylor and product-formula components while retaining an error bound of \(0.02\) relative to the specified finite quadrature.

\section{Overview}
\label{sec:overview}

\paragraph{One caller, two providers}
A Boolean oracle gives coherent access to a predicate \(f\colon\{0,1\}^n\to\{0,1\}\). On a basis state with address \(x\) and answer bit \(b\), the reversible query maps \(\ket{x,b}\) to \(\ket{x,b\oplus f(x)}\), where \(\oplus\) denotes XOR. The action on superpositions follows by linearity~\cite{beals2001quantum}. If the answer qubit is prepared in \(\ket{-}=(\ket{0}-\ket{1})/\sqrt{2}\), the query leaves that qubit unchanged and multiplies the amplitude at \(x\) by \((-1)^{f(x)}\). The induced operation on the address is a \emph{phase oracle}.

Consider a caller that passes the ordered address view \([5,2]\) to a phase slot \(f\) under a control on a separate wire \(g\). The slot marks a single address, with the first wire of its argument corresponding to the low bit. The caller's proof uses this interface to conclude that its guarded operation contributes a phase of \(-1\) exactly when \(g=1\) and the physical address \((q_5,q_2)\) is \((1,0)\). The client can later bind a gate-only implementation or a QRAM implementation. The gate provider flips phase at the specified bit pattern. The QRAM provider loads \([x=1]\) into a private flag, applies \(Z\), and repeats the load to clear the flag. This example exposes the boundary conditions of the linking theorem. The Grover case in Section~\ref{sec:grover} uses the same provider pattern in a search circuit.

\begin{figure}[!htbp]
\centering\small
\begin{tabularx}{\linewidth}{@{}lXX@{}}
\toprule
Binding obligation & Correct binding & Observable consequence of a change \\
\midrule
Ordered actuals & Pass \(q_5\) as the low bit and \(q_2\) as the high bit. & Swapping the view moves the phase from physical pattern \((1,0)\) to \((0,1)\). \\
Lexical snapshot & Forward the caller's \(\mathsf{marked}\) bank to the provider's \(\mathsf{table}\). & A same-shape bank with entry \(1\) at address \(2\) marks a different address. \\
Public matrix & Prove the signed phase action, with \(+1\) on unmarked addresses. & Multiplying the provider by \(-1\) reverses the relative phase of the two values of \(g\). \\
Private workspace & Prove the flag is zero at return for every public input. & Omitting the final read leaves the flag entangled with an address superposition. \\
\bottomrule
\end{tabularx}
\caption{Four independent obligations for a guarded phase slot. The graph checks ordered actuals and handles. Provider proofs establish the matrix and workspace obligations.}
\Description{A table of four binding errors, with examples of how an incorrect wire order, bank, phase, or flag affects the caller.}
\label{fig:overview}
\end{figure}

\paragraph{What the caller can observe}
Fix the address at a marked basis state and prepare the guard in \(\ket{+}\). The specified provider leaves the guard in \(\ket{-}\); its globally negated variant leaves it in \(\ket{+}\). A Hadamard followed by a guard readout distinguishes the two with certainty. For the private flag, remove the final read and start the address in \((\ket{x_{\mathrm{marked}}}+\ket{x_{\mathrm{other}}})/\sqrt{2}\) with guard one. The provider produces
\[
 \bigl(-\ket{x_{\mathrm{marked}}}\ket{1}
       +\ket{x_{\mathrm{other}}}\ket{0}\bigr)/\sqrt{2}.
\]
The two flag states are orthogonal, so ignoring the flag destroys the address coherence the caller needs. More generally the return condition is an equation on the entire input subspace with zero private wires. Linearity transfers it from basis inputs to superpositions, including inputs entangled with an external reference.

\paragraph{Which equalities the checker uses}
The verifier compares the decoded open graph with a view of the submitted bound graph that hides the selected slot bodies. It first reconstructs approved bank forwarding from the binding plan. This comparison catches the first two changes in Figure~\ref{fig:overview}, even though each preserves the signature. A different implementation of the same phase oracle can pass this check because the selected body is opaque at that boundary. The phase and reset obligations are discharged by separate implementation evidence. Exact linking then yields the matrix used by the caller proof at the ordered actuals and forwarded snapshot. If two callers share a provider, the proof for its lexical interface can be reused with each caller's own bank.

This design makes the obligations of a proposed binding explicit. Programs can change how slots are implemented and introduce helpers inside a provider. An optimization of the retained caller, even cancellation of two adjacent inverse operations, can change the graph equality. Accepting such a rewrite would require independent equivalence evidence for the caller. The structural decision avoids solving that equivalence problem during every binding. Likewise, the general theorem takes clean exit as a semantic contract. For the read--phase--read provider, a circuit-specific proof discharges it on every input. Section~\ref{sec:evaluation} reports rejection tests that exercise these distinctions.

\paragraph{Where approximation enters}
An ordinary unitary slot can bound the error of its entire public action. Such a bound passes through placement and coherent control and contributes once per invocation in the caller. A block-encoding slot exposes signal wires and gives a guarantee only on a selected input and output block. A call can move amplitude out of that block and a later call can return it. For example, a signal-only \(X\) and \(Y\) both have zero \(\ket{0}\)-to-\(\ket{0}\) block, whereas \(XX\) and \(XY\) have different blocks. A projected replacement therefore needs a fact about how its caller uses the slots. In LCHS, orthogonal selector labels ensure that the output is a weighted sum of separately projected branches. This denotation supports a weighted error rule without treating projection as an ordinary clean module return.

\section{Open programs and late binding}
\label{sec:open-programs}

Figure~\ref{fig:open-language} gives the register-level fragment used by the linking theory and shows what remains open and what binding produces. A slot is a named module declaration with a signature, a declared oracle role, and an absent body. The signature fixes the ordered quantum registers and any bank formals already exposed by the slot. A call to a slot uses the same syntax and interface checks as a call to an implemented module. Oracq provides a Python builder and a serialized intermediate representation (\rir{}) for this fragment~\cite{chen2026oracq}. The builder fixes widths, gate parameters, and repetition counts. The linking theorems operate over decoded graphs.

\begin{figure}[!htbp]
\centering
\setlength{\fboxsep}{7pt}%
\begin{lrbox}{\openlanguagebox}
\begin{minipage}{\dimexpr\linewidth-2\fboxsep-2\fboxrule\relax}
\small
\setlength{\abovedisplayskip}{4pt}%
\setlength{\belowdisplayskip}{4pt}%
\setlength{\abovedisplayshortskip}{4pt}%
\setlength{\belowdisplayshortskip}{4pt}%
\textbf{Register-level syntax}
\[
\begin{gathered}
 p ::= \langle e,\mathcal M\rangle
 \qquad D ::= \langle\bar r,\bar b,\bar\ell,d\rangle \\
 d ::= \mathsf{open}\mid c
 \qquad c ::= \epsilon\mid i\,c
 \qquad v ::= \overline{(r,s,h)}\\
 \begin{aligned}
 i&::=g(v)\mid\mathsf{gphase}(\theta)\mid\mathsf{xor}(v_1,v_2)
       \mid\mathsf{swap}(v_1,v_2)\mid\mathsf{add}(v,k)\\
  &\quad\mid\mathsf{load}(b,v_a,v_d)\mid\mathsf{call}(m,\bar v,\bar b)
       \mid\mathsf{repeat}(k,c)\mid\mathsf{ctrl}(v,k,c)\mid\mathsf{adj}(c)
 \end{aligned}
\end{gathered}
\]
A program has entry \(e\) and module table \(\mathcal M\).
A definition \(D\) lists public registers \(\bar r\), formal banks \(\bar b\), private registers \(\bar\ell\), and body \(d\). The grammar omits module attributes.
\par\smallskip\hrule\smallskip
\textbf{Slots and binding}
\[
\begin{aligned}
 \mathsf{Slots}(p) &= \{f \mid \mathcal M(f).d=\mathsf{open}\} \\
 \beta &= \{ f_j\mapsto (I_j,\eta_j) \}_j,
   \qquad \operatorname{dom}(\beta)\subseteq\mathsf{Slots}(p) \\
 p_\beta &= \mathsf{bind}(p,\beta)
\end{aligned}
\]
Here \(I_j\) implements slot \(f_j\), and \(\eta_j\) assigns any additional bank parameters of \(I_j\) to banks at the entry. The entry is open if a slot is reachable from it. The bound entry is closed if every reachable definition has a body.
\par\smallskip\hrule\smallskip
\textbf{A Boolean oracle bound to a truth table}
\[
 \beta=\{f\mapsto(\mathit{bit\_table},\{t\mapsto\mathit{truth}\})\}.
\]
\noindent
\begin{minipage}[t]{0.43\linewidth}
\textbf{Open RIR \(p\)}
\begin{lstlisting}[basicstyle=\ttfamily\footnotesize,aboveskip=2pt,belowskip=0pt]
slot f(x:n, y:1)
module dj(x:n, y:1)
  X(y); H(y); H(x)
  call f(x,y)
  H(x)
\end{lstlisting}
\end{minipage}\hfill
\begin{minipage}[t]{0.55\linewidth}
\textbf{Bound RIR \(p_\beta\)}
\begin{lstlisting}[basicstyle=\ttfamily\footnotesize,aboveskip=2pt,belowskip=0pt]
module dj(x:n, y:1 | truth:Q(n,1))
  X(y); H(y); H(x)
  call f(x,y | truth)
  H(x)
module f(x:n, y:1 | b:Q(n,1))
  call bit_table(x,y | b)
module bit_table(x:n, y:1 | t:Q(n,1))
  load(t,x,y)
\end{lstlisting}
\end{minipage}
\end{minipage}
\end{lrbox}
\fbox{\usebox{\openlanguagebox}}
\caption{Open RIR and late binding. The two programs are shown in abbreviated module notation. \(x:n\) declares an \(n\)-bit register and \(Q(n,1)\) is a QRAM bank with \(n\) address bits and one data bit. The slot \(f\) declares the Boolean XOR oracle interface. In the bound program, \(b\) is the formal parameter through which \(f\) receives \(\mathit{truth}\). The bank contents are supplied in a separate snapshot.}
\Description{A boxed grammar, an open-slot and binding-map definition, and a small Deutsch--Jozsa example. The same caller invokes f before and after binding. After binding, the caller passes the truth bank to f, which forwards it to a module that performs a QRAM read.}
\label{fig:open-language}
\end{figure}

In Figure~\ref{fig:open-language}, \(g\) ranges over \(H\), \(X\), \(Z\), \(R_y(\theta)\), and the single-qubit phase gate, with each gate broadcast over the view. Quantum registers carry a width and a storage kind, and views retain their kind. Banks carry an address width and a data width. An empty instruction block \(\epsilon\) implements the identity. The fragment uses fixed read-only snapshots and places measurement and classical readout outside the circuit.

Binding checks the selected implementations against the slot interfaces. The entry in Figure~\ref{fig:open-language} still calls \(f\), whose new body delegates to the table lookup. The additional bank parameter makes the source of the lookup explicit. Other slots can remain open after binding. Section~\ref{sec:link} relates an open and a closed program using an independently supplied specification of how banks are forwarded.

\section{The scoped abstract machine and its semantics}
\label{sec:resolution}

Figure~\ref{fig:scoped-machine} defines the execution of a resolved program. Each module has a relative frame with its public wires followed by private wires. Resolution maps each register to its ordered wires and concatenates the selected slices in span order. The first wire is the least significant bit. Thus \([5,2]\) encodes a two-bit word with wire~5 low. A resolved call records the callee, ordered quantum views, and typed bank arguments. Appendix~\ref{app:frontend} describes the resolution checks and their comparison with Oracq's reference execution.

The verifier can expand implemented calls into an instruction sequence in the entry frame or preserve the module graph. In the latter case each module is interpreted in its own frame. Different calls can apply the same module on different wires and with different banks.

\begin{figure}[!htbp]
\centering
\setlength{\fboxsep}{7pt}%
\fbox{\begin{minipage}{\dimexpr\linewidth-2\fboxsep-2\fboxrule\relax}
\small
\setlength{\abovedisplayskip}{4pt}%
\setlength{\belowdisplayskip}{4pt}%
\setlength{\abovedisplayshortskip}{4pt}%
\setlength{\belowdisplayshortskip}{4pt}%
\renewcommand{\arraystretch}{1.18}%
\textbf{Wire resolution}
\[
 \mathsf{wires}_{L_m}(v)=\mathop{\Vert}_{(r,s,h)\in v} L_m(r)[s:s+h]
\]
\par\smallskip\hrule\smallskip
\textbf{Scoped machine}
\[
\begin{aligned}
 \text{Configuration} &\quad \langle m,\Sigma_m,w,A,K\rangle\\
 \text{Call continuation} &\quad R=\langle m,\Sigma_m,w,A,a,C\rangle\\
 \text{Start at a concrete entry} &\quad \langle e,\Sigma_e,w_e,I,[]\rangle
\end{aligned}
\]
A configuration has module \(m\), lexical snapshot \(\Sigma_m\), remaining instructions \(w\), accumulated matrix \(A\), and continuation stack \(K\). For a resolved call \(q=(m',\bar v,\bar b)\), let \(a\) concatenate its wire lists and let \(\Sigma'\) be the forwarded snapshot. The context \(C[-]\) applies the call's enclosing guards and adjoints, and acts as the identity for a call without either.
\smallskip

\begin{tabular}{@{}l@{\quad}l@{}}
Primitive &
  \(\langle m,\Sigma_m,iw,A,K\rangle \to \langle m,\Sigma_m,w,MA,K\rangle\)
  \quad if \(i\) has action \(M\) \\
Call &
  \(\langle m,\Sigma_m,C[q]w,A,K\rangle \to
    \langle m',\Sigma',w_{m'},I,R::K\rangle\) \\
Return &
  \(\langle m',\Sigma',\epsilon,U,R::K\rangle \to
    \langle m,\Sigma_m,w,C[\mathsf{place}_a(V)]A,K\rangle\) \\
Open call &
  \(\langle m,\Sigma_m,C[q]w,A,K\rangle \to
    \langle m,\Sigma_m,w,C[\mathsf{place}_a(\rho(m',\Sigma'))]A,K\rangle\) \\
Halt &
  \(\langle e,\Sigma_e,\epsilon,U,[]\rangle \to \mathsf{done}(V)\)
\end{tabular}
\smallskip

The Call rule enters a concrete body \(w_{m'}\). If the slot is open, the Open call rule uses its public action from the oracle environment \(\rho\). Return and Halt require the clean-exit rule below.
\par\smallskip\hrule\smallskip
\textbf{Instruction and call semantics}
\smallskip

\begin{tabularx}{\linewidth}{@{}l@{\quad}X@{}}
Gates and phase &
  \(g(v)\) applies \(g\) on the selected wires. A global phase has action \(e^{i\theta}I\). \\
Word operations &
  XOR updates the target by the source bits. Swap exchanges two words.
  Addition updates \(z[v]\) to \((z[v]+k)\bmod 2^{|v|}\). \\
QRAM read &
  \(z[v_d] \gets z[v_d]\oplus\Sigma_m(b)(z[v_a])\) on basis states. \\
Control &
  \(\ctrl_P(U)=PUP+(I-P)\), where \(P\) projects onto the guard sector. \\
Adjoint and repeat &
  \(U^\dagger\) and \(U^k\), respectively. \\
Bank forwarding &
  \(\Sigma'(f_j)=\Sigma_m(b_j)\) for the callee's \(j\)th bank formal \(f_j\). \\
Clean exit &
  \(U\) unitary and \((FF^\dagger)UF=UF\), where \(F\ket{x}=\ket{x,0^\ell}\)
  appends the private wires. The public action is \(V=F^\dagger UF\). \\
Wire placement &
  \(\mathsf{place}_a(V)=P_a^\dagger(V\otimes I)P_a\), where \(P_a\) moves the
  bits at \(a\), in order, ahead of the spectator bits.
\end{tabularx}
\end{minipage}}
\caption{Execution semantics of the scoped Oracq abstract machine. The call and return rules specify how a slot or concrete module acts inside the caller.}
\Description{A boxed definition of ordered wire resolution and the machine configurations and transitions. The lower table gives instruction actions, bank forwarding, placement, and clean public returns.}
\label{fig:scoped-machine}
\end{figure}

Let \(B_n=\{0,1\}^n\). The accumulator describes the action on every input in \(\mathbb C^{B_n}\) in the active frame. An instruction with matrix \(M\) multiplies the accumulator on the left, so \(\operatorname{compose}([M_1,\ldots,M_t])=M_t\cdots M_1\). A missing slot action or a failed return check has no successful execution.

\subsection{Semantics of instructions and call boundaries}
\label{sec:scoped}

Primitive instructions leave spectator wires fixed. A serialized angle denotes its exact decimal value in radians. A bank of shape \((a,d)\) supplies a function \(B_a\to B_d\). With disjoint address and data views, the QRAM read in Figure~\ref{fig:scoped-machine} is an involution and extends linearly to superpositions. XOR and swap act on raw bit patterns. Constant addition requires an unsigned view.

The control rule keeps the phase of \(U\) relative to the disabled sector. Its unitarity follows when \(U\) is unitary and preserves the guard sectors. Adjoint takes the conjugate transpose and reverses the factor order of a sequential product. A block repeated zero times has action \(I\).

The clean-exit equation is equivalent to zero amplitudes from every input with zero workspace to every basis state with nonzero workspace. It implies \(UF=FV\) and, for unitary \(U\), gives a unitary \(V\). Lean proves that clean exits are preserved by adjunction and composition. Clean exit also commutes with control when the public guard agrees with the guard on the zero-workspace subspace and guard preservation holds. These properties justify releasing a module's private wires at return and reusing them on subsequent calls.

\FloatBarrier
\section{Validating and linking a binding}
\label{sec:link}
\label{sec:scoped-link}

The checker takes two independently decoded programs and a bank-forwarding plan \(\Phi\). A plan belongs to the caller's verification task. Its entries are indexed by retained caller and direct callee, so two callers can forward different banks to a shared definition. Each entry lists the callee's bank arguments in its bound signature order. An item \(\mathsf{old}(i)\) forwards original actual \(i\) and \(\mathsf{capture}(b)\) names a bank in the caller's lexical frame. This section defines the structural relation the checker decides and the semantic condition under which that relation transfers a proof.

\subsection{Resolution and structural validation}

Write \(\Gamma_m=(L_m,\mathcal B_m)\) for a module's register layout and lexical bank signature, and \(\operatorname{resolve}(p)\Downarrow G\) for successful scoped resolution. Resolution checks the register and bank signatures and traverses the reachable calls in dependency order. If \(m'\) expects registers of types \(\bar\tau\) and banks of shapes \(\bar\sigma\), the call rule instantiates its ordered interface as follows. Here \(\bar a\) is a list of ordered wire lists, \(\bar h\) is a list of typed bank handles and \(\operatorname{nd}\) means no duplicates.
\[
\frac{
  \Gamma_m\vdash\bar v:\bar\tau\rightsquigarrow\bar a
  \qquad \operatorname{nd}(\operatorname{concat}(\bar a))
  \qquad \Gamma_m\vdash\bar b:\bar\sigma\rightsquigarrow\bar h
}{
  \Gamma_m\vdash\mathsf{call}(m',\bar v,\bar b)
    \rightsquigarrow\mathsf{query}(m',\bar a,\bar h)
}
\tag{Call resolution}
\]
The view premise concatenates each sequence of in-bounds spans and checks its declared width and register kind. The bank premise checks shape and retains the actual lexical handle. Bank arguments may alias. The snapshot effect is given by \(\Sigma'(f_i)=\Sigma_m(h_i)\), with the shape equality used to interpret the value at formal \(f_i\). Thus shape equality permits forwarding, and the handle determines \emph{which} bank a read observes.

\paragraph{Control and adjoint resolution}
A resolution context carries an ordered guard stack \(\bar P\) and an adjoint bit \(d\). In the rules below, \(R_{\bar P,d}\) is resolution in the same lexical layout and \(\circ\) is sequencing of resolved words.
\[
\begin{aligned}
 R_{\bar P,d}(\mathsf{adj}(c)) &= R_{\bar P,1-d}(c),\\
 R_{\bar P,d}(\mathsf{ctrl}_{Q}(c)) &= R_{\bar P\mathbin{+}[Q],d}(c),\\
 R_{\bar P,0}(c_1c_2) &= R_{\bar P,0}(c_1)\circ R_{\bar P,0}(c_2),\\
 R_{\bar P,1}(c_1c_2) &= R_{\bar P,1}(c_2)\circ R_{\bar P,1}(c_1).
\end{aligned}
\]
Each emitted operation \(A\) is wrapped in the guards in stack order, with \(A^\dagger\) at the center when \(d=1\). The checker requires a nonempty, in-range guard, disjoint guard views, and no overlap between protected wires and call actuals. For a coherent read, the address may overlap protected wires and the destination may not. A controlled or adjoint call must have the required capability throughout its callee subtree. Repetitions have statically known counts and are expanded in the current context. These equations describe the supported cases of the resolver. The matrix laws below justify distribution over sequences. The named-source comparison in Appendix~\ref{app:frontend} checks a subset of this normalization at the decoded RIR boundary.

\paragraph{Reconstruction}
For a retained caller and a direct call with original actuals \(\bar h\), resolve the plan list by
\[
  \bigl[\operatorname{actual}(s,\bar h)\bigr]_{s\in\Phi(m,m')},
  \quad
  \operatorname{actual}(\mathsf{old}(i),\bar h)=h_i,
  \quad
  \operatorname{actual}(\mathsf{capture}(b),\bar h)=b .
\]
If the plan has no entry, use \(\bar h\). A present entry uses each original position exactly once. For a retained callee the plan must preserve the original bank formals and may introduce only distinct new formals. Each retained caller keeps its quantum signature and private register types and may add only approved captured banks to its own bank signature. Recursion under control and adjoint changes only the bank list at a query. It retains the ordered quantum views and both modifiers.

Given resolved graphs \(G=\operatorname{resolve}(p)\) and \(G'=\operatorname{resolve}(p_\beta)\), let \(S\) be the reachable bodyless definitions of \(G\). The partial function \(\operatorname{rebuild}_\Phi(G,G')\) checks the entry, interfaces and plan above and requires \(G'\) to be closed. It starts from the layout and dependency order of \(G'\). For retained callers it replaces each body with the body from \(G\) after bank reconstruction. For \(S\) it uses the selected implementation interfaces and leaves their bodies absent. Other definitions keep their bodies and layouts, so an implementation can introduce private helpers. The structural judgment is
\[
\frac{
 \operatorname{resolve}(p)\Downarrow G
 \quad \operatorname{resolve}(p_\beta)\Downarrow G'
 \quad \operatorname{rebuild}_\Phi(G,G')\Downarrow E
 \quad E=\operatorname{hide}_S(G')
}{
 p\;\bowtie_{\Phi}\;p_\beta\;\triangleright E
}
\tag{Binding validation}
\]
Here \(\operatorname{hide}_S\) erases just the bodies at \(S\). The graph equality compares the reconstructed instruction sequences, including guards, adjoints, ordered actuals and typed banks. It does not compare the selected bodies with the oracles' mathematical contracts.

\begin{lemma}[Structural decision and evidence]
\label{lem:validation}
The executable scoped checker accepts exactly when the binding-validation judgment has a derivation. Acceptance constructs evidence for its four premises.
\end{lemma}

Each successful resolution and reconstruction contributes its defining equation to the evidence. The final equality is decidable. Conversely, each premise determines the corresponding successful branch of the checker. This is completeness for the structural judgment above. Equality is tested after resolution, so the call comparison already absorbs spelling differences in register slices and the supported relocation of control and adjoint modifiers. An optimization that cancels two gates in the retained word can still fail. The preservation result below is stated on the values returned by decoding the inputs.

An implementation can interpose a validator for a small set of rewrites. The LCHS assembly word comparison gives a concrete example. It erases exact zero-angle rotations and phases and their guarded or adjointed forms. The proof shows that each erased operation is identity as a full matrix, which also accounts for coherent control. That normalization belongs to the assembly check. To admit such rewrites at the general binding boundary, one would additionally need a proof that the rewrite preserves every retained body with its calls left abstract, including its ordered views and lexical bank arguments. In this design the current general checker decides a structural relation, and a transformation pass can establish that extra equivalence separately.

\subsection{Matrix invariants at a call}

\begin{lemma}[Ordered placement and guards]
\label{lem:placement}
Let \(a\) be an injective ordered wire view. Placement preserves products, adjoints, the identity and unitarity. For any guard projector \(P\) on wires disjoint from \(a\), \([P,\mathsf{place}_a(U)]=0\). A coherent read also commutes with a guard on its address wires if its write view is disjoint from the guard.
\end{lemma}

For basis rows \(r\) and columns \(c\), let \(r|_a\) be the tuple of bits in the order of \(a\) and \(\bar a\) the spectator wires. The placed entry is \(U_{r|_a,c|_a}\) if \(r|_{\bar a}=c|_{\bar a}\), and zero otherwise. Splitting a basis index into the ordered tuple and spectators is a bijection. Under this bijection placement is reindexing of \(U\otimes I\), which proves the algebraic laws. Nonzero entries agree on every spectator, which proves guard preservation. A coherent read fixes its address and all bits outside its write view, which gives the claimed read law. For example, the views \([5,2]\) and \([2,5]\) can both satisfy the disjointness check and place different matrices: the low bit of the callee is different. The validator therefore compares \(a\) as an ordered tuple.

Write \(\mathsf{Pres}_P(U)\) for the requirement that the cross-sector matrix entries are zero, equivalently \([P,U]=0\). Guard preservation is closed under products and adjoints. Under this condition the controlled action of a unitary is unitary, and
\[
 \ctrl_P(UV)=\ctrl_P(U)\ctrl_P(V),
 \qquad
 \ctrl_P(U^\dagger)=\ctrl_P(U)^\dagger .
\]
The product law follows by expanding with \(P(I-P)=0\) and using guard preservation. The adjoint equality also holds without guard preservation. The product law justifies distributing control over a word; adjoint instead reverses its order and conjugates each factor. These are the matrix invariants used when interpreting the normalized bodies.

\begin{lemma}[Closure of clean exits]
\label{lem:clean}
Let \(F\) be a module's zero-workspace frame and \(V=F^\dagger U F\). If the unitary \(U\) has a clean exit, then \(UF=FV\), \(V\) is unitary and \(U^\dagger F=FV^\dagger\). If \(U_iF=FV_i\) for \(i=1,2\), then \((U_2U_1)F=F(V_2V_1)\). For a diagonal guard \(P\) with restriction \(Q=F^\dagger P F\), one has \(\ctrl_P(U)F=F\ctrl_Q(V)\). If \(U\) preserves the guard, this controlled action is also unitary.
\end{lemma}

The frame satisfies \(F^\dagger F=I\). Thus \(UF=FV\) yields \(V^\dagger V=I\), and finite dimension gives its inverse \(V^\dagger\). Multiplying the frame equation by the inverses gives the adjoint equation. Substituting it twice for successive bodies gives the product law. Diagonal guards satisfy \(PF=FQ\). Substituting this equation into \(PUP+(I-P)\) proves the control law. The clean-exit condition is needed on the entire public input subspace. A branch used for block encoding can leave its \emph{public} signal wires nonzero, which the caller later projects onto zero. A module's \emph{private} wires must satisfy the clean-exit condition at return.

\paragraph{Discharging a clean exit for QRAM}
The verifier has a constructive rule for the phase provider in Section~\ref{sec:overview}. Its premise matches the resolved body to two identical reads of the same ordered addresses and destination with a \(Z\) on a private data bit between them. Write \(L_b\) for the first read and \(x\) for the ordered public address. On basis inputs,
\[
   L_b Z L_b\ket{x,0}
        =(-1)^{\Sigma(b)(x)}\ket{x,0}.
\]
The read permutation is an involution on the full frame, so conjugating a diagonal phase by it is diagonal. Hence every transition from a zero-private input to an outside basis state vanishes. The provider proof then identifies the surviving phase from the forwarded bank, and the full public-input clean-exit property follows by linearity. This rule works at any address width. It discharges clean exit for the QRAM provider in Section~\ref{sec:grover}; arbitrary providers supply their own proof.

\subsection{From an accepted binding to caller correctness}

Let \(\rho(f,\Sigma_f)\) be the specified public action of slot \(f\), a partial function of its lexical snapshot. For every \(f\in S\), the slot contract states that whenever \(\rho(f,\Sigma_f)\) succeeds, the chosen concrete module (with its private dependencies) runs on \(\Sigma_f\) and performs that public action after a clean unitary exit. We relate partial denotations by successful-result preservation: \(A\sqsubseteq B\) means \(A=\mathsf{ok}(M)\) implies \(B=\mathsf{ok}(M)\) for every result \(M\).

\begin{theorem}[Binding validation for scoped programs]
\label{thm:scoped-link}
Suppose \(p\bowtie_\Phi p_\beta\triangleright E\) and every selected implementation fulfills its slot contract. For every entry snapshot \(\Sigma\), the abstract evaluation of \(E\) with slots interpreted by \(\rho\) is related by \(\sqsubseteq\) to the scoped evaluation of the decoded bound program \(p_\beta\). The two successful results are equal as complex matrices.
\end{theorem}

\begin{proof}
Graph equality identifies \(E\) with the bound graph whose slot bodies are opaque. Evaluate both graphs in their common dependency order. The induction invariant relates their module environments pointwise in lexical snapshots: the paired entries have the same name, public width and bank formals, and their meanings are related by \(\sqsubseteq\). The slot-contract hypothesis supplies the invariant at a slot. At a retained call, both sides look up the same callee interface. Ordered placement and snapshot forwarding are identical, so each successful abstract callee matrix is also produced by the concrete callee. Its placed matrix agrees on the full caller frame, including every spectator. Applying identical control and adjoint contexts preserves that equality. Pointwise agreement on successful instruction results gives agreement on the word product. A successful unitary and clean-exit check therefore yields the same compressed public matrix. This extends the environment invariant, which applied to the entry establishes the conclusion.
\end{proof}

The invariant relates functions of the lexical snapshot. If two calls to a shared definition use distinct banks, the same module proof applies to each forwarded snapshot. Equality includes phase and justifies substituting inside a coherent control. The structural checker can compare two graphs before obtaining the slot proofs, and the theorem does not establish the contracts simply from matching signatures.

\subsection{Approximate public actions}
\label{sec:quantitative-link}

An exact slot contract can be relaxed to a phase-sensitive error budget on the public matrix. We use the induced Euclidean operator norm. A budget can depend on the lexical snapshot at a call, since distinct banks passed to one definition can carry different data. The layout and forwarding checks are the same as in exact linking. The reference and replacement must each give a clean unitary exit, since a small public matrix error alone does not establish this invariant.

\begin{lemma}[Propagation at the call boundary]
\label{lem:call-distance}
Let \(U,V\) be comparable matrices, \(a\) be an injective ordered view, and \(P\) be a diagonal guard projector. Ordered placement preserves the norm of the difference. Adjoint also preserves it, and coherent control with the same guard does not increase it. Compression between common isometric frames \(F_L,F_R\) is nonexpansive:
\[
\begin{aligned}
 \opnorm{\mathsf{place}_a(U)-\mathsf{place}_a(V)}&=\opnorm{U-V},\\
 \opnorm{U^\dagger-V^\dagger}&=\opnorm{U-V},\\
 \opnorm{\ctrl_P(U)-\ctrl_P(V)}&\le\opnorm{U-V},\\
 \opnorm{F_L^\dagger U F_R-F_L^\dagger V F_R}&\le\opnorm{U-V}.
\end{aligned}
\]
\end{lemma}

Placement tensors the difference with the identity on the spectator wires and changes basis by a permutation. The active sectors of the two controlled matrices differ by \(P(U-V)P\), since their inactive identities cancel. The two frame maps are contractions. For two contractions \(U_1,U_2\) and their replacements \(V_1,V_2\), the sequential law follows from
\[
 U_2U_1-V_2V_1=(U_2-V_2)U_1+V_2(U_1-V_1).
\]
The two product terms have norms at most \(\opnorm{U_2-V_2}\) and \(\opnorm{U_1-V_1}\).

\begin{theorem}[Quantitative linking at a scoped body]
\label{thm:scoped-distance}
Take a retained resolved body at a common snapshot. Suppose its reference and replacement evaluations have clean unitary exits, and their individual instruction matrices are contractions. For each call occurrence \(i\), let \(\varepsilon_i\) bound the difference between its two callee public matrices at the snapshot forwarded by that occurrence. Then the two public results \(V,V'\) of the body satisfy
\[
                    \opnorm{V-V'}\le \sum_{i\text{ a call occurrence}}\varepsilon_i .
\]
\end{theorem}

In the proof, primitive instructions agree because their parameters and snapshot agree. For each call, placement lifts the bound at its actual view. The control and adjoint cases apply Lemma~\ref{lem:call-distance} to the underlying instruction. Telescoping the word product adds the resulting bounds, and compression to the common zero-workspace frame preserves the bound.

\begin{theorem}[Quantitative linking of an accepted binding]
\label{thm:scoped-approx}
For \(p\bowtie_\Phi p_\beta\triangleright E\), pair the entries of the abstract environment of \(E\) and the concrete environment of the bound graph. A quantitative evidence table assigns each pair a snapshot-dependent public budget. At slots this is a provider contract. At a retained module \(m\) with contraction evidence as above, extend the table with
\[
 \Delta_m(\Sigma)
   =\sum_{i\in\mathsf{calls}(m)}\Delta_{f_i}
                         \bigl(\mathsf{forward}_i(\Sigma)\bigr).
\]
If both entry interpretations succeed, the matrix error is at most the budget of the entry at the supplied snapshot.
\end{theorem}

The Lean evidence table aligns names, widths and bank formals in both environments. Extending it in dependency order uses Theorem~\ref{thm:scoped-distance} for each retained body. The binding evidence identifies the abstract environment with the reconstructed caller's environment, and the resolved bound graph determines the concrete environment. Matching entries translates the public budget to the results of the two decoded programs. A shared callee used twice contributes twice to the bound even though its contract is proved once. This rule compares successful interpretations. The exact rule additionally transfers successful evaluation.

For example, a module \(h\) can forward formal bank \(b\) to an approximate slot \(f\) under a control. Its budget is \(\Delta_h(\Sigma)=\Delta_f(\Sigma_f)\), where the rebased snapshot satisfies \(\Sigma_f(\mathsf{table})=\Sigma(b)\). An entry that calls \(h\) with two same-shape banks, once under an adjoint on a reordered view, incurs one charge at each of the two forwarded snapshots. Reordering which caller wires are passed changes the matrices but leaves the placement norm law unchanged. Reordering the actuals between the open and bound artifacts would fail the graph check.

\paragraph{Checking a phase provider end to end}
The command-line checker accepts a serialized caller with a phase slot \(f\), an independently serialized binding, a decimal reference angle \(a\), and an error limit. Each round calls \(f\) coherently under a guard on a separate wire and then calls \(\mathsf{adj}(f)\). A globally phased provider therefore changes a relative phase in the caller. The checker reads the implementation's angle \(b\) from the decoded provider. The contract on its whole public matrix is
\[
  \opnorm{e^{ia}I-e^{ib}I}=2|\sin((a-b)/2)|\le |a-b|.
\]
The checker validates both graphs and computes the rational bound \(2k|a-b|\) for \(k\) rounds. The proof constructs the evidence table by extending the checked slot contract to the retained caller. Theorem~\ref{thm:scoped-approx} then bounds the distance between the abstract caller and the actual decoded bound program. This check requires only comparisons on the resolved structure and arithmetic on the extracted decimals. Section~\ref{sec:evaluation} reports the accepted and rejected inputs.

\subsection{Projected contracts}
\label{sec:projected-link}

A block contract bounds a specified compression of a slot's public action. An arbitrary caller can mix the specified subspace with its complement between calls, as in Section~\ref{sec:overview}. To reuse only block contracts, we require a separate fact about the enclosing caller. One reusable rule applies when its compressed action is a weighted sum of the independently compressed slot matrices.

\begin{lemma}[Weighted block replacement]
\label{lem:weighted-replacement}
Suppose the output of a linked caller, including its normalization, satisfies \(B=\sum_j \widehat c_j B_j\), where \(B_j\) is the normalized block of its \(j\)-th provider. If the specified output is \(\sum_j c_j V_j\) and \(\opnorm{V_j-B_j}\le\varepsilon_j\), then
\[
 \opnorm{\sum_j c_j V_j-B}\le
   \sum_j\bigl(|c_j|\varepsilon_j+|c_j-\widehat c_j|\opnorm{B_j}\bigr).
\]
\end{lemma}

The premises separate the caller's denotation from its providers' contracts. The proof expands the difference using \(c_j(V_j-B_j)+(c_j-\widehat c_j)B_j\), where scalar multiplication is understood. The inequality is valid for any finite selector set, logical block dimension, and complex coefficients. It makes no claim about the unprojected matrix. For a select--prepare caller, the premise about \(B\) follows from placement on the complement of a complete selector guard, orthogonality of those guards, and the actual preparation amplitudes. Section~\ref{sec:lchs} proves that the submitted LCHS bindings satisfy this premise.

\section{Executable verification and certified generation}
\label{sec:interface}

\begin{figure}[t]
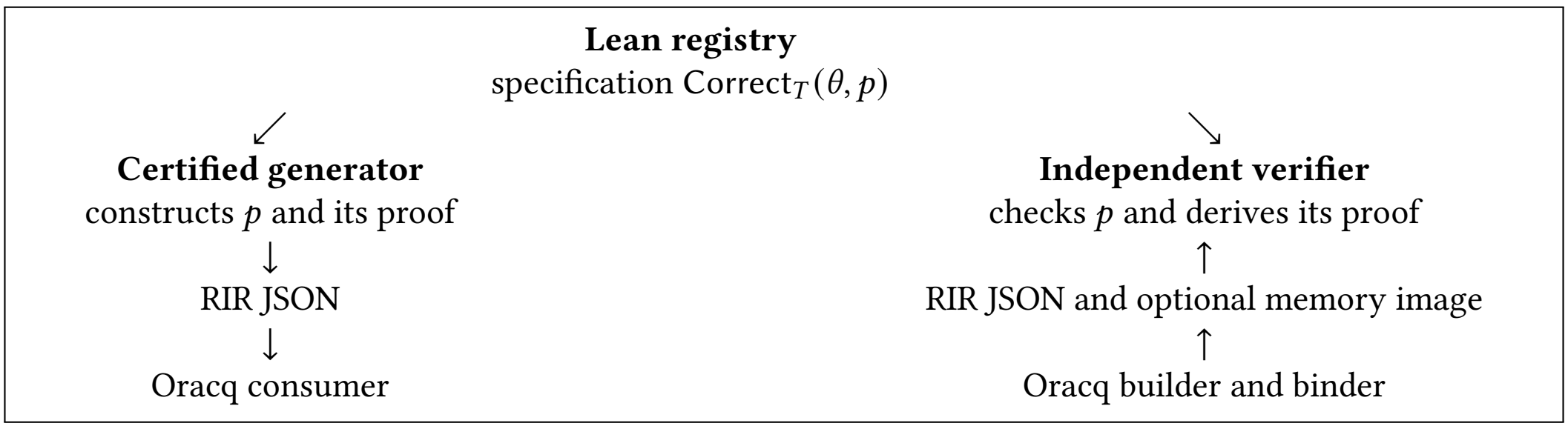

\centering
\fbox{\begin{minipage}{\dimexpr\linewidth-2\fboxsep-2\fboxrule\relax}
\centering\small
\begin{tabular}{@{}c@{\quad}c@{\quad}c@{}}
 & \textbf{Lean registry} & \\
 & specification \(\mathsf{Correct}_T(\theta,p)\) & \\
 \(\swarrow\) & & \(\searrow\) \\
 \textbf{Certified generator} & & \textbf{Independent verifier} \\
 constructs \(p\) and its proof & & checks \(p\) and derives its proof \\
 \(\downarrow\) & & \(\uparrow\) \\
 RIR JSON & & RIR JSON and optional memory image \\
 \(\downarrow\) & & \(\uparrow\) \\
 Oracq consumer & & Oracq builder and binder
\end{tabular}
\end{minipage}}
\caption{Generation and independent checking share a Lean specification. The generator serializes its program for Oracq. Either a generated program or a separately constructed program can be submitted to the verifier.}
\Description{A two-branch workflow from a Lean registry. The left branch constructs a program and proof and exports RIR JSON. The right branch decodes input from the Oracq builder or binder and checks the decoded program against the same specification.}
\label{fig:lean-interface}
\end{figure}

Figure~\ref{fig:lean-interface} shows the two routes through the system. In the generation route the client selects a registered Lean specification and supplies parameters such as an address width. In the verification route an independent producer supplies a program. The verifier decodes it and checks the hypotheses needed to apply the same specification. The binding route also takes the open caller, the candidate bound program, and a snapshot when the program uses QRAM. The Python binder produces the candidate bound program. The checker independently reconstructs the approved structure as in Section~\ref{sec:link}.

\subsection{Registering an executable specification}

An entry for theorem name \(T\) contains a parameter type \(\Theta_T\), a parser for the supported parameters, and a predicate \(\mathsf{Correct}_T:\Theta_T\to\mathsf{Program}\to\mathsf{Prop}\). The Lean types of the two entry points are
\[
\begin{aligned}
 \mathsf{generate}_T(\theta)&:\{p:\mathsf{Program}\mid\mathsf{Correct}_T(\theta,p)\},\\
 \mathsf{check}_T(\theta,p)&:\mathsf{Except}\;\mathsf{Error}\;
             \{u:\mathsf{Unit}\mid\mathsf{Correct}_T(\theta,p)\}.
\end{aligned}
\]
The registry dispatches on the fully qualified theorem name. Each entry supplies executable code and a proof for the declared predicate. The open verifier can accept a program from another producer if it resolves to the certified word or graph and meets the additional checks for that specification. Each entry also supplies its own bound checker, including a snapshot decoder and implementation checks when needed. Its soundness uses the corresponding case-specific Lean theorems.

\paragraph{Example of the proof flow}
For Deutsch--Jozsa, \(\Theta_T\) consists of positive widths. Generation constructs a register layout and a single XOR slot. The open checker independently resolves its argument \(p\) and establishes the required word equality at width \(n\). Both routes use the same theorem from that equality to the oracle-parametric final-state formula. For the bound route the verifier also certifies a constant, one-bit or dense-table oracle and checks the Boolean promise. For Grover, the registry parameter is one of four marked addresses. The open check establishes a scoped graph equality and checks any declared address. Its mathematical theorem fixes the phase oracle at the requested address. The QRAM-bound check additionally validates that the snapshot has the required table and proves the private flag is restored. Section~\ref{sec:case-studies} gives the readout guarantees. Appendix~\ref{app:implementation} gives example commands.

\paragraph{Proofs and serialized artifacts}
The generator constructs a dependent pair in Lean. The CLI emits the program component as JSON. The independent verifier reconstructs its own evidence on decoding this JSON or another producer's JSON. Its result is an acceptance report, with the logical theorem stated over the decoded Lean value. The scoped structural judgment also allows a binding to be checked before the selected implementations' semantic contracts have been supplied. Algorithm-specific checks compose this evidence with the relevant contracts. Deutsch--Jozsa and Grover use the registry to expose both directions. LCHS uses an independent verifier with a separately checked selector and branch certificates. Its structural binding check is the same, and its additional proof establishes the projected premise of Lemma~\ref{lem:weighted-replacement}.

\subsection{Trusted computing base}
\label{sec:tcb}

\paragraph{Mathematical statements and executable results}
The core theorems are about decoded programs, typed memories, the exact denotation of decimal angles, and the scoped interpreter. For these theorems the kernel checks the proofs and their dependencies, including the matrix, QRAM and algorithm models. The Python producer and binder can supply arbitrary artifacts. They do not need to be correct for a successful checked statement about the decoded values. A claim about the meaning of the original Python program requires a separate frontend correspondence. The named-RIR comparison described in Appendix~\ref{app:frontend} begins at decoded RIR.

The CLI passes bytes through the Lean JSON parser and the RIR decoder. These components determine which typed value is checked, and the theorems do not specify their full byte-level behavior. Interpreting a CLI success report therefore also relies on the decoder and on executing the checker and reporting its result faithfully. In particular, interpreting a shipped executable requires trusting the toolchain that runs it. This execution boundary is distinct from the use of native evaluation as evidence inside a Lean proof.

\paragraph{Axioms and native evaluation}
We run \texttt{\#print axioms} on the headline soundness theorems, including binding validation, certified Grover generation, and the linked LCHS error bound. Table~\ref{tab:axioms} records which declarations rely on compiled decisions inside their proofs. The Grover generator checks the four resolved caller instances with kernel-checked reduction. In Lean 4.35, \texttt{native\_decide} introduces a dedicated axiom for each compiled decision~\cite{lean-reference-axioms}. The two fixed comparisons for the original nine-wire standalone Taylor certificate use these axioms. The linked LCHS checker instead derives bounds from hypotheses recorded by executable certificates. The separate seven-wire Taylor branch uses kernel reduction for its fixed comparisons.

\begin{table}[t]
\centering
\caption{Proof dependencies from \texttt{\#print axioms}. Standard axioms are \texttt{propext}, \texttt{Classical.choice}, and \texttt{Quot.sound}.}
\label{tab:axioms}
\small
\begin{tabularx}{\linewidth}{@{}Xl@{}}
\toprule
Audited claims & Additional native decisions \\
\midrule
Scoped structural decision and semantic linking & 0 \\
Scoped public error propagation and weighted block replacement & 0 \\
End-to-end quantitative phase binding & 0 \\
Guarded SELECT and select--prepare denotation & 0 \\
Deutsch--Jozsa generation and bound verification & 0 \\
Grover independent bound verification and parameterized readout & 0 \\
Grover certified generation & 0 \\
Linked LCHS structure-to-matrix proof and finite-quadrature error & 0 \\
Wide Grover bound probability and wide LCHS block encoding & 0 \\
Standalone seven-wire Taylor branch & 0 \\
Standalone nine-wire Taylor circuit & 2 \\
\bottomrule
\end{tabularx}
\end{table}

\section{Case studies}
\label{sec:case-studies}

\subsection{Exact replacement across Grover widths and iterations}
\label{sec:grover}

Grover search isolates the phase and memory obligations of a late-bound oracle~\cite{grover1996search}. Let \(n>0\) be the address width, let \(a\) be the marked address, and let \(O_a\ket{x}=(-1)^{[x=a]}\ket{x}\). The caller first prepares \(\ket{s}=H^{\otimes n}\ket{0^n}\), then repeats a phase-oracle call followed by diffusion. With \(D=2\ket{s}\!\langle s\rvert-I\) and iteration count \(k\), the open caller denotes
\[
       (D O_a)^k H^{\otimes n}.
\]
To implement the zero-state reflection, the circuit applies \(Z\) to wire \(i\) conditioned on all preceding wires being zero. Its phase is \(-1\) precisely on states whose first set bit is at position \(i\). Every nonzero basis state has exactly one first set bit, while the zero state satisfies none of those conditions. The product therefore has diagonal \(+1\) on zero and \(-1\) on every other address. This construction fixes the phase of the reflection exactly. Each broadcast Hadamard layer implements \(H^{\otimes n}\) because each wire contributes exactly one tensor factor. Conjugating the reflection by these layers gives \(D\). Lean proves these identities for resolved words, then composes the \(k\) rounds using the same oracle contract at every occurrence.

The state from \(\ket{0^n}\) agrees with the singleton-search state formalized in Qatlib~\cite{qatlib}. With \(\theta_n=\arcsin(2^{-n/2})\), its success probability is
\[
        \Pr[\mathrm{readout}=a]=\sin^2\!\left((2k+1)\theta_n\right).
\]
We check two independently bound providers at every supported width, marking the address with only the highest bit set. The gate provider applies the final prefix-controlled \(Z\) used in the reflection construction. The QRAM provider loads the certified singleton table into a zero private flag, applies \(Z\), and repeats the load to clear the flag. The checker validates the graph boundary, both levels of bank forwarding, and the implementation together with its snapshot. The width-parametric caller theorem then supplies the result for the decoded bound program. At \(k\) iterations the providers use zero and \(2k\) QRAM reads, respectively.

The registered two-bit generator uses a circuit with an equivalent diffusion word and certifies exact preparation at \(k=1\). Its independent verifier handles all four marked addresses with a QRAM provider and address \(3\) with a controlled \(Z\) provider. The width-parameterized binding verifier is a separate entry point over submitted programs. Its theorem starts from checked resolution, while a width-parameterized registry entry would also have to prove that its generator always produces a program with that resolution. Appendix~\ref{app:cases} gives the commands for checking submitted wide programs.

\subsection{Approximate replacement in one LCHS caller}
\label{sec:lchs}

Linear combination of Hamiltonian simulation (LCHS) expresses dissipative dynamics as a weighted integral of unitary evolutions~\cite{an2023lchs}. We use the noncommuting model \(L=\operatorname{diag}(3/2,1/2)\), \(H=X/4\), and \(t=1/8\). The four-node left-endpoint rule uses \(k_j=-1+j/2\), \(c_j=(1/2)/(\pi(1+k_j^2))\), and \(V_j=e^{-it(H+k_jL)}\).

We generate one open Oracq program with four branch slots. Its coefficient preparation, the ordered controlled calls, and inverse preparation are retained across bindings. Each slot accepts one target wire and a common seven-wire signal register, with normalization two. One implementation uses a second-order Taylor polynomial. Another uses the first-order product formula \(e^{-itX/4}e^{-it k_j Z/2}e^{-it k_j}\)~\cite{childs2021trotter}. Each adapter embeds the implementation's own signal wires and uses a separate rotation flag to match the slot's normalization. The Taylor and product implementations can therefore be substituted one node at a time. Both bound callers have ten wires including the two selector wires.

For a branch implementation \(U_j\), let \(B_j=2J^\dagger U_j J\), where \(J\) embeds the target with zero signal. Let \(B\) be the assembled success block scaled by the caller's declared normalization.

\paragraph{A compositional law for SELECT and PREPARE}
The algebraic step can be proved without evaluating the enclosing circuit. Let \(P\) be the exact preparation matrix on the selector register and \(a_j=\langle j|P|0\rangle\). In the selector-first basis, a call guarded by label \(j\) acts as \(U_j\) on that sector and as identity on all others. A word that contains exactly one such call per label realizes \(\operatorname{diag}(U_j)_j\). The proof restricts the product to each sector. Disjointness of the branch view and selector wires justifies this interpretation of a placed call, including noncontiguous views (Lemma~\ref{lem:placement}).

\begin{lemma}[Projected denotation of a select--prepare word]
\label{lem:select-prepare}
Suppose the caller has the matrix of a PREPARE, complete SELECT, and inverse PREPARE in the above basis, with common branch normalization \(\alpha>0\). Let \(B_j=\alpha J^\dagger U_j J\) for the shared zero-signal frame \(J\), and let \(B\) be the assembled block scaled by \(\lambda\). Then \(B=\sum_j\widehat c_j B_j\), with \(\widehat c_j=\lambda|a_j|^2/\alpha\).
\end{lemma}

The proof compresses block-diagonal SELECT against the prepared selector on both sides. Each diagonal sector contributes \(\overline{a_j}a_j\). Inserting the signal frame gives the claimed weighted sum. For an accepted binding, the remaining proof obligation is to identify the matrix and frames of the submitted program with those of this composition.

\paragraph{Checking the denotation of the submitted assembly}
The checker independently resolves the bound entry and each branch provider. It uses the open selector word \(p\), the bound branch words \(u_j\), the selector view \(s=[1,2]\), and the branch view \(b=[0,3,4,5,6,7,8,9]\). It constructs the following expected word in physical wire order:
\[
 \mathsf{place}_s(p)\ ;\ 
   \Bigl[\mathsf{ctrl}_{s=j}(\mathsf{place}_b(u_j))\Bigr]_{j=0}^{3}\ ;\ 
   \mathsf{adj}(\mathsf{place}_s(p)).
\]
Here placement renames every wire and nested guard in a local word. The outer control is distributed over the placed instructions, and adjoint reverses the preparation list and adjoints each instruction. The submitted root word must match this expected word, after erasing only rotations or phases with exactly zero angle and their controlled or adjoint forms. This comparison includes the entry wrapper. For example, the emitted initialization \(R_y(0)\) is accepted as identity. Changing it to \(R_y(0.125)\) in both the open and bound artifacts preserves their structural linkage and fails the assembly check.

\begin{lemma}[Denotation of the checked assembly]
\label{lem:lchs-artifact}
If the structural word comparison succeeds, the decoded root matrix in selector-first coordinates is
\[
 (P^\dagger\otimes I)\operatorname{diag}(U_0,U_1,U_2,U_3)(P\otimes I),
\]
where \(P\) and \(U_j\) are the exact matrices of the checked selector and local branch words. Its physical success frame embeds the data bit at wire zero with every other wire zero.
\end{lemma}

The proof first relocates each elementary gate and its nested guards in the ordered view. The placement law for products then relocates entire words. The selector wires and branch wires partition the caller's wires. Each branch operation preserves the selector, so its sequence of controlled gates equals control of the composed branch. The four guards cover the selector values exactly once. Finally, the reindexing that moves the selector to the front sends the physical zero-signal frame to the zero-selector tag of the branch frame. Applying Lemma~\ref{lem:select-prepare} identifies the projected matrix of the actual decoded artifact with the weighted branch blocks.

The executable checker estimates columns of the individual branches and the selector. The open selector certificate is reused for all three bindings. It bounds each \(|c_j-\widehat c_j|\) by \(10^{-12}\), and the unitary branch matrices give \(\opnorm{B_j}\le 2\). With \(\varepsilon_j=0.015\) and \(\sum_j|c_j|\le1\), Lemma~\ref{lem:weighted-replacement} gives a finite-quadrature error bound \(0.02\) for all Taylor, all product formula, and the mixed binding. The verifier rejects changes to the selector, exchanges of two node implementations, and perturbations of a normalization adapter. Section~\ref{sec:lchs-evaluation} compares the cost with a checker that evaluates the two logical columns of the complete bound circuit.

\paragraph{Widening the coupled target}
Let \(m\) be the number of target wires, and let \(C_m\) apply a controlled \(X\) from the original data wire to each of the \(m-1\) additional wires. Define the lifted action \(\mathcal L_m(A)=C_m^\dagger(A\otimes I)C_m\) on the logical register. The widened problem has
\[
 H_m=\mathcal L_m(H)=\tfrac14 X^{\otimes m},\qquad
 L_m=\mathcal L_m(L)=I+\tfrac12(Z\otimes I).
\]
Thus the Hamiltonian flips all target bits simultaneously and fails to commute with the dissipative part. The family is unitarily equivalent to a one-site model tensored with identity. This gives an experiment in scaling the enclosing circuit with unchanged branch contracts. The branch certification problem stays one-site at every tested width.

The wide open caller precedes the same four-slot assembly by \(C_m\) on its public data wires and follows it by \(C_m^\dagger\). The checker validates the open and bound graph boundary and then compares the full resolved gate word to this composition. The proof identifies the resolved controlled flips with a basis permutation, and shows that permutation preserves the physical success frame and restricts to \(C_m\) on its logical coordinates. Compression therefore gives \(\mathcal L_m(B)\) for the scaled block \(B\) of the inner assembly. The lift preserves addition, multiplication, scalars, and the matrix exponential, so \(\mathcal L_m(\sum_j c_j V_j)\) is the four-node quadrature for \(H_m,L_m\). Since \(\mathcal L_m\) preserves the norm of any difference, all three wide bindings inherit the \(0.02\) bound. The widened comparison traverses symbolic gate words and does not evaluate columns at the full circuit width.

\subsection{Parameterized Deutsch--Jozsa}
\label{sec:dj}

The Deutsch--Jozsa caller in Figure~\ref{fig:open-language} has an address register $x$ of any positive width $n$, an answer bit $y$, and the slot $f$. Its logical interface is the standard XOR oracle
\[
  U_f\ket{x,y}=\ket{x,y\mathbin{\oplus}f(x)} .
\]
The caller prepares $\ket{0^n}\ket{-}$, applies $H$ to all address bits, invokes $f$ once, and applies $H$ to all address bits again~\cite{deutsch1992rapid}. The final address amplitude at $z$ is the Walsh coefficient
\[
  a_f(z)=2^{-n}\sum_{x\in\{0,1\}^n}(-1)^{f(x)+x\cdot z}.
\]
The answer remains $\ket{-}$. A constant function yields address zero with probability one. A balanced function has $a_f(0)=0$. These statements are parameterized by the positive width and the promise on $f$.

A concrete implementation may use a constant circuit, an XOR from one selected address bit, or a coherent read from a dense truth table. The table is supplied separately as a memory snapshot. All three choices satisfy the same logical XOR interface when its Boolean function is interpreted accordingly. A table with an invalid promise must be rejected even if its signature and access pattern match. At \(n=1\), exchanging the arguments \(x\) and \(y\) preserves their types and changes which wire holds the address. Resolution makes this difference explicit.

The parameterized proof supplies the shared specification in Section~\ref{sec:interface}. The benchmark varies the width for a symbolic bit oracle and the number of entries for a dense QRAM bank. These runs exercise two different scaling costs.

\section{Evaluation}
\label{sec:evaluation}
\label{sec:evidence}

Table~\ref{tab:evidence} summarizes the guarantees. We also test controlled and adjoint calls that pass ordered views and distinct memory banks, then compare the cost of checking expanded and shared module representations.

\begin{table}[t]
\caption{Verification results for the experimental programs.}
\label{tab:evidence}
\centering
\small
\begin{tabularx}{\linewidth}{@{}>{\raggedright\arraybackslash}p{0.18\linewidth}*{2}{>{\raggedright\arraybackslash}X}@{}}
\toprule
Case & Established result & Scope \\
\midrule
Deutsch--Jozsa & One oracle call and exact readout under the Boolean promise. & Every positive width. Bound examples include constant, bit, and table oracles. \\
Controlled and adjoint calls & Exact matrix preservation by Theorem~\ref{thm:scoped-link}. & Successful validation and the theorems' matrix contracts. \\
Grover & A width- and iteration-parameterized caller proof transfers the search probability to both bindings. & One marked address at each width, with a gate provider and a dense QRAM provider. \\
LCHS & The same open caller accepts both methods and a mixed binding within \(0.02\) of the finite quadrature. & Four slots and a coupled target family. Numerical evaluation stops at the selector and branch boundaries. \\
\bottomrule
\end{tabularx}
\end{table}

\paragraph{Repeated Grover calls}
We check both Grover bindings with \(k=4{,}096\). This input expands to \(24{,}577\) root instructions, including \(4{,}096\) slot calls. The check transfers the symbolic readout theorem and exercises repeated use of the same provider. The input has two address wires throughout, so this experiment varies call count only. At \(k=2,4,17\), independent reference simulation gives the probabilities \(1/4,1,1/4\), respectively.

\paragraph{Growing the Grover address register}
For the width-parameterized caller we check both bindings at \(n=1,3,5,8,12,16\), using different positive iteration counts. At \(n=16\), the QRAM interface reads a 65\,536-entry singleton bank and allocates one private flag. At that width and \(k=4\), the median is 3.00\,s with gates and 5.85\,s with QRAM including snapshot checks. The gate checker also runs at \(n=128\), \(k=2\), with a median of 3.10\,s. It checks the prefix guards symbolically. Simulation at \(n=1,3,5\) agrees with the probability formula for both providers. A snapshot that moves the mark and a provider that omits the cleanup read both fail verification.

\paragraph{Control, adjoint, and memory}
Two further experiments combine coherent controls, adjoint calls, ordered slices, and multiple QRAM banks. The verifier accepts the intended bindings and rejects changes to wire order, bank identity, or the specified bank forwarding. These mutations preserve wire counts or bank shapes, so signature matching alone would accept them. The version with private workspace also agrees with a reference evaluation of complex amplitudes on all 32 address and control branches. Appendix~\ref{app:cases} describes both circuits and their rejection tests.

\paragraph{Quantitative linking of a phase slot}
We use \(a=0.25\) and \(b=0.2509765625\), so each call has a full-matrix budget \(1/1024\). The checked bounds for \(1,8,64,512\) rounds are \(1/512,1/64,1/8,1\), respectively. The median CLI times for these inputs are 2.85, 2.83, 2.82 and 2.86 seconds, over three runs after a warm-up. At 64 rounds the checker rejects a limit of \(1/16\). It also rejects a submitted program whose round count disagrees with the parameter. This experiment exercises the quantitative scoped theorem with both call modifiers. The LCHS experiments exercise the projected rule.

\paragraph{Verification time}
Table~\ref{tab:runtime} reports median process times for one compiled verifier, including initialization, decoding, verification and reporting. Artifact generation is completed before invoking the verifier. The no-input \texttt{list} command measures the startup baseline. A debugger sample places the verifier in an eagerly evaluated finite check in an imported module before command dispatch. This shared initialization accounts for most of the wall time of small inputs. These measurements include symbolic verification, a dense QRAM table with \(2^{16}\) words, and a structural comparison of 256 calls at one placement or at 256 distinct placements. Appendix~\ref{app:timings} gives the protocol.

\begin{table}[t]
\caption{Median elapsed time for verification at selected input sizes.}
\label{tab:runtime}
\centering
\small
\begin{tabular}{@{}llr@{}}
\toprule
Check & Input & Time (s) \\
\midrule
Initialization baseline & No input program & 2.846 \\
Deutsch--Jozsa, bit oracle & 1 address bit & 2.848 \\
Deutsch--Jozsa, bit oracle & 128 address bits & 2.866 \\
Deutsch--Jozsa, QRAM table & 16 words & 2.867 \\
Deutsch--Jozsa, QRAM table & 65\,536 words & 3.125 \\
Controlled-phase link & 256 calls, 1 placement & 2.940 \\
Controlled-phase link & 256 calls, 256 placements & 2.955 \\
\bottomrule
\end{tabular}
\end{table}

\subsection{Cost of structural LCHS assembly}
\label{sec:lchs-evaluation}

The numerical assembly baseline uses the same open program, selector certificate and branch checks. It estimates two columns of the complete bound word to compare its compressed matrix to the weighted blocks. The structural checker instead compares the submitted word to the word reconstructed from the component traces. Both establish the same \(0.02\) bound on the finite quadrature. For a total width \(N\), the baseline maintains a vector of \(2^N\) amplitudes for each input column. After resolution and lowering, the structural checker traverses the instruction lists and their guard operands. The branch checks in both implementations still maintain vectors of the branch widths. In the linked examples the largest numerically evaluated state vector falls from \(2^{10}=1{,}024\) entries to \(2^8=256\) entries. The selector itself has four entries.

Table~\ref{tab:lchs-runtime} records a separate paired comparison between the numerical and structural assembly implementations. It reports medians over five runs per executable after one warm-up. Within each row, both versions run on the same logical CPU of the Xeon Gold 5118 host, and the order alternates across repetitions. The timings cover the command from startup through parsing, checking, and reporting. They include every selector and branch certificate required for a single binding. The structural checker is between 3.1 and 4.3 times as fast on these inputs. The remaining time includes the numerical branch checks.

\begin{table}[t]
\centering
\caption{Linked LCHS verification using whole-circuit columns or structural assembly. Times are median seconds for a single binding. The bound words each use ten wires.}
\label{tab:lchs-runtime}
\small
\begin{tabular}{@{}lrrr@{}}
\toprule
Binding & Resolved instructions & Whole-circuit columns & Structural assembly \\
\midrule
Taylor in all slots & 159 & 47.10 & 10.93 \\
Product formula in all slots & 43 & 15.34 & 4.93 \\
Mixed binding & 54 & 18.62 & 5.63 \\
\bottomrule
\end{tabular}
\end{table}

\paragraph{Growing the LCHS target register}
We check all Taylor, all product formula, and mixed bindings for the coupled family at 2, 4, 8, and 16 data wires. Table~\ref{tab:lchs-wide} counts resolved instructions including the controlled flips in the outer conjugation. Every accepted binding satisfies the same \(0.02\) bound against its own four-node quadrature. At 25 total wires an explicit circuit column would have \(2^{25}\) amplitudes. The largest vector evaluated by the checker has 256 entries at every width, because the numerical certificates are for the original branch interfaces. Removing the inverse fanout from both submitted programs preserves their structural binding and fails the wide assembly check. Changing the declared outer normalization is also rejected.

\begin{table}[t]
\caption{Widened LCHS circuits checked with three bindings of a single caller at each width. Time is the median for checking all three in one invocation.}
\label{tab:lchs-wide}
\centering\small
\begin{tabular}{@{}rrrrrr@{}}
\toprule
Data wires & Total wires & Taylor gates & Product gates & Mixed gates & Time (s) \\
\midrule
2 & 11 & 161 & 45 & 56 & 15.20 \\
4 & 13 & 165 & 49 & 60 & 15.25 \\
8 & 17 & 173 & 57 & 68 & 15.45 \\
16 & 25 & 189 & 73 & 84 & 15.42 \\
\bottomrule
\end{tabular}
\end{table}

\paragraph{Symbolic checking at larger register widths}
We also check a symbolic balanced oracle at 128, 512, and 1\,024 address wires. The provider XORs the last address bit into the answer bit. The checker proves the balance condition and derives the nonzero readout of the submitted bound caller using the symbolic theorem of Section~\ref{sec:dj}. The median times in the same batch as Table~\ref{tab:runtime} are 2.87, 3.11, and 3.94 seconds. The largest circuit uses 1\,025 wires and 2\,053 gates under the counting convention of Appendix~\ref{app:cases}. An explicit state vector has \(2^{1025}\) entries. This check examines the resolved program and verifies the symbolic oracle contract.

\paragraph{Shared implementations}
Table~\ref{tab:shared-runtime} isolates the cost of preserving a callee definition across callers. The circuit width stays at two while each additional level doubles the number of primitive instructions in the expanded representation and adds one scoped definition. At depth 18, the expanded checker completes in 52.40 seconds and the scoped checker completes in 2.88 seconds. At depth 20, the scoped checker accepts 23 definitions, while the expanded executable reports an allocation failure under the 4\,GiB address-space limit. Both commands construct their usual reports, which differ in size. This experiment measures structural validation and reporting; proving the provider's semantic contract is a separate operation.

\begin{table}[t]
\caption{Binding validation on a shared call graph. Times are median seconds. At depth 20 the expanded count is derived from the graph, and the expanded executable reports an allocation failure under the 4\,GiB address-space limit.}
\label{tab:shared-runtime}
\centering
\small
\begin{tabular}{@{}rrrrr@{}}
\toprule
Depth & Expanded instructions & Scoped definitions & Expanded & Scoped \\
\midrule
0 & 16 & 3 & 2.852 & 2.860 \\
4 & 256 & 7 & 2.960 & 2.873 \\
8 & 4\,096 & 11 & 3.020 & 2.867 \\
10 & 16\,384 & 13 & 3.303 & 3.038 \\
14 & 262\,144 & 17 & 5.672 & 2.958 \\
18 & 4\,194\,304 & 21 & 52.404 & 2.878 \\
20 & 16\,777\,216 & 23 & Allocation failure & 2.866 \\
\bottomrule
\end{tabular}
\end{table}

\subsection{Formalization size and build}
\label{sec:formalization-size}

Table~\ref{tab:lean-size} counts first-party Lean modules reachable from the proofs for the language, both registered specifications, and the case studies, including those in the appendices. It counts physical source lines and explicit declaration commands, excluding the command-line driver, generated declarations, and imported libraries. The counted sources contain no \texttt{sorry}, \texttt{admit}, or explicit \texttt{axiom} declarations. The theorem audit in Table~\ref{tab:axioms} additionally checks the transitive dependencies of the headline proofs, including the axioms introduced by native evaluation.

We measured two incremental \texttt{lake build} runs with Lean~4.35.0 on an x86-64 Intel Xeon Gold 5118 host. Adding the quantitative phase module to the proof build took 145 seconds. Rebuilding the revised Grover generation proof and its dependents then took 21 seconds. Both runs used compiled public dependencies and reused the unchanged first-party modules.

\begin{table}[t]
\centering
\caption{First-party Lean source for the paper. Definitions include \texttt{def}, \texttt{abbrev}, \texttt{structure}, and \texttt{inductive} declarations.}
\label{tab:lean-size}
\small
\begin{tabular}{@{}lrrrr@{}}
\toprule
Component & Modules & Lines & Theorems/lemmas & Definitions \\
\midrule
RIR, semantics, and linking & 43 & 8\,527 & 336 & 324 \\
Specifications and generation & 3 & 356 & 6 & 26 \\
Case-study verification & 21 & 4\,775 & 220 & 226 \\
\midrule
Total & 67 & 13\,658 & 562 & 576 \\
\bottomrule
\end{tabular}
\end{table}

\section{Related work}

\paragraph{Verified quantum compilers and generators}
SQIR and VOQC provide a Coq semantics for circuits and verify transformations against complex matrices~\cite{hietala2021voqc}. VQO verifies compilation of arithmetic and geometric oracle implementations through OQASM~\cite{li2022vqo}. QSV combines a verified translation for a small quantum language with validation of state-preparation programs~\cite{li2026qsv}. Qbricks combines parameterized circuit construction with deductive verification~\cite{chareton2021qbricks}. Parameterized Verification of Quantum Circuits verifies relational properties of circuit families using weighted tree automata~\cite{abdulla2026parameterized}. These systems motivate explicit theorems about executable circuit families. Our formulation exposes the ordered wires and banks of each call to proofs about the caller. The Deutsch--Jozsa and Grover generators and verifiers each share a specification for generated and independently supplied programs.

PC-QEC uses Lean to check symbolic certificates for families of quantum error-correction circuits under specified error and readout contracts~\cite{ye2026pcqec}. Lai et al. verify Hamiltonian-simulation circuits by extracting the ancilla block and the implemented quantum signal processing polynomial~\cite{lai2026qsp}. Their verifier checks the block-encoding oracle once and uses its abstract interface to check the enclosing transformation. Their layer-specific argument avoids expanding oracle calls in the quantitative verifier. Our scoped theorem handles substitution at a decoded call site with arbitrary ordered wire arguments, lexical bank forwarding, and phase under control. The LCHS experiment keeps the coefficient preparation and call structure fixed. The checker compares the decoded assembly to a reconstructed gate word and derives the block error from the component certificates.

\paragraph{Logics, control, and approximation}
CoqQ gives a foundational logic for quantum programs with register reasoning~\cite{zhou2023coqq}. QbC provides refinement rules for correctness by construction in a quantum while language~\cite{peduri2025qbc}. QafnyC embeds quantum verification into Dafny~\cite{cheng2025qafnyc}. Qunity develops compositional quantum control and efficient compilation~\cite{mints2025qunity}. Venev et al.\ synthesize uncomputation and adjoints modularly~\cite{venev2024uncomputation}. A synthesized implementation can meet our clean-exit obligation if it has the required proof about the entire public input space. SAQR-QC tracks weights described by local projections and observables across quantum circuits~\cite{yu2026saqr}. Our numerical check bounds an operator-norm error on a zero-signal projection. Its companion linking theorem tests the call boundary that makes separately justified implementations substitutable.

\paragraph{Linking and memory}
Work on verified separate compilation connects separately compiled components through semantic preservation and linking obligations~\cite{kang2016separate}. In our workflow the source and bound graphs are already in one register-level semantics, and the validator checks the correspondence between their retained calls. This concentrates the proof at the binding boundary. Optimizations that change retained words require a separate equivalence guarantee, particularly where control can observe phase. Quantum random-access stored-program machines model a broader memory and program architecture~\cite{wang2023qram}. Our memory model uses typed read-only banks, a fixed snapshot, and coherent XOR reads. Bank arguments determine which memory a callee sees.

\section{Limitations and conclusion}
\label{sec:scope}

\paragraph{Language fragment}
Our programs have fixed-width quantum registers, acyclic calls, and statically known repetition counts. The interpreter uses read-only QRAM snapshots and coherent XOR loads. Measurement, classical adaptation, dynamic loops, and writable quantum memory are outside the formalized fragment. Clean exit is a restriction on a module's complete public input subspace, so intentionally returning garbage or keeping unreturned workspace entangled with that subspace falls outside the exact slot contract. Supporting those features would require an effectful semantics for the state passed across a call and a corresponding contextual contract.

\paragraph{Verification scope and cost}
The scoped theorems are parameterized by widths, banks and module graphs. The executable structural checker has explicit preprocessing limits, including bounds on nested blocks and repeat expansion. Deutsch--Jozsa and the Grover caller are parameterized by positive address width. The wide Grover binding check uses a fixed singleton pattern at each width to compare its two providers. Its gate path checks a symbolic phase word, while the QRAM path must read all \(2^n\) entries of the explicitly supplied bank image.

The linked LCHS artifact uses four quadrature nodes. Its \(0.02\) bound concerns that finite sum. Appendix~\ref{app:lchs} gives a \(0.48\) bound between the base quadrature and continuous evolution, including cutoff and discretization. It also gives a \(0.47\) continuous-evolution bound for the original standalone Taylor and product-formula circuits. The wide family tests the cost of checking an enclosing circuit as its register grows. Its representation preserves the original one-site branch certification problem, so the experiment keeps both the branch dimension evaluated and the number of quadrature nodes fixed. The numerical checker retains its limits of ten wires and 512 resolved instructions for such evidence. Scaling the branch problem to an irreducible multi-site Hamiltonian or supplying more quadrature nodes would exercise different costs.

\paragraph{Portability and semantic boundary}
The theory is phrased over a register-level graph with injective ordered views, abstract calls, lexical read-only bank arguments, and a zero-workspace convention. A different front end could use the same proof obligations by translating its calls and supplying a correspondence between its denotation and that graph. In particular the correspondence must preserve phase in the presence of control and preserve the distinction between the bank's shape and the snapshot it designates. Oracq provides the implementation studied here. The front-end and export checks in Appendix~\ref{app:frontend} validate specific translations at their stated boundaries. Physical-device behavior is outside the model.

The register-level format lets a caller be serialized while its oracle slots remain open. Once a binding is supplied, \ov{} checks the resulting program. The linking theorems transfer the caller's proof under the implementation contracts, so the same proof can serve multiple deployments of an oracle-based algorithm.

\paragraph{Use of generative AI} OpenAI Codex assisted in drafting this manuscript and developing the Lean formalization and Python experiments.

\section*{Code Availability}

The \ov{} source code is available under the MIT License. The version accompanying this paper is \texttt{v0.1.0} at \url{https://github.com/quantum-atlas-ai/OracqVerify/releases/tag/v0.1.0}. It contains the Lean formalization, the generator and validation CLI, and scripts for the case studies and benchmarking workloads, with a pinned toolchain and public dependencies. Appendix~\ref{app:implementation} describes the reproduction workflow.

\bibliographystyle{ACM-Reference-Format}
\bibliography{refs-arxiv}
\appendix
\section{Implementation and reproduction}
\label{app:implementation}
\label{app:frontend}

\paragraph{Reproduction workflow}
The Lean build checks the formalization and builds the \texttt{oracq-verify} executable. The Python scripts use a pinned Oracq revision to construct and bind RIR programs and supply QRAM snapshots where required. The regression suite in \texttt{scripts/check.sh} exercises the registered generators and independent checks, including rejection of incorrect bindings. The scripts \texttt{check\_grover.py}, \texttt{check\_wide\_cases.py}, and \texttt{check\_replacements.py} construct the Grover and LCHS artifacts and run their respective validation commands. For the timing workloads described in Appendix~\ref{app:timings}, run \texttt{bench\_verification.py} in \texttt{--paper} and \texttt{--scoped} modes with \texttt{--repetitions 3} in each run.

\paragraph{Program construction and resolution}
The Python \texttt{Builder} assembles nested instruction blocks, the entry, and its dependencies. It orders the register and bank arguments of a call according to the callee's signature. The Python binding operation fills selected declarations with forwarding definitions and adds bank parameters to the modules that need them. For the program in Figure~\ref{fig:open-language}, the corresponding Python call is:
\begin{lstlisting}
p_bound = bind(p, {"f": Binding(bit_table, {"t": "truth"})})
\end{lstlisting}
The two serialized programs are decoded separately by the verifier. The Lean theorems concern the decoded programs and the relation checked between them.

Resolution checks slice bounds, widths, storage kinds, and disjointness of quantum arguments. Banks may be passed to more than one formal parameter when their shapes agree. Resolution checks the capabilities required under control and adjoint and protects the guard wires from writes. A QRAM read may use those wires as its address when its data view is disjoint. In the scoped representation, resolution keeps one body per reachable module. It expands repetitions, reverses the relevant instruction order under adjoint, and records controls and adjoints on individual instructions. The peak number of private wires is the module's own private width plus the largest such count of any of its children, with zero child contribution at a leaf.

\paragraph{Lean declarations}
\texttt{linkWord\_sound}, \texttt{SourceLink.sound}, and \texttt{ScopedSourceLink.sound} give Theorems~\ref{thm:link}, \ref{thm:artifact-link}, and \ref{thm:scoped-link}, respectively. Lemma~\ref{lem:validation} is \texttt{verifyScopedSourceLink\_iff}.

The following declarations give the body bound, its lifting to the shared environment, the bound on an accepted artifact, the weighted block bound, and the SELECT compression law, respectively.
\begin{lstlisting}
ScopedModule.distance_le
ScopedErrorEvidence.extendBody
ScopedSourceLink.distance_le
weighted_blocks_error
selectPrepare_compress
\end{lstlisting}
The width-parameterized Grover diffusion and the two wide artifact guarantees use the following declarations. The last declaration is the end-to-end quantitative phase guarantee.
\begin{lstlisting}
WideGrover.reflectionWord_matrix
WideGrover.Certificate.probability
WideLCHS.liftBlock_quadrature
WideLCHS.Certificate.sound
ApproximatePhase.Certificate.sound
\end{lstlisting}

\paragraph{Command-line interface}
A generator is selected by its Lean theorem name. For Deutsch--Jozsa with three address bits, the following commands generate and verify the open program and verify a supplied QRAM-bound version. The last command takes the separately bound program and the memory contents as inputs.
\begin{lstlisting}
theorem=OracqVerify.Generators.DeutschJozsa.generate_correct
oracq-verify generate "$theorem" 3 > dj.open.json
oracq-verify verify "$theorem" dj.open.json 3
oracq-verify verify-bound "$theorem" \
    dj.open.json dj.qram.bound.json 3 memory.json
\end{lstlisting}
Grover uses the theorem \texttt{OracqVerify.Generators.Grover.generate\_correct} with a marked address from zero through three. Its proof uses kernel-checked reduction to show that each constructed caller resolves to the verified graph and records the supplied address. The commands \texttt{check-link} and \texttt{check-scoped-link} validate bindings using the expanded and scoped representations, respectively. They take the open and bound programs and, where needed, the specification of bank forwarding. Case-specific commands also check the selected oracle or algorithm contract.

\paragraph{Comparison with the reference execution}
For programs without local workspace, the regression tests compare the instructions expanded by Oracq with those resolved in Lean. The comparison includes ordered wire views, guard values, adjoint direction, decimal angles, and bank arguments. Tests with local workspace also compare nested calls under control and adjoint. These tests exercise the path from Python construction and serialization to the verified representation.

\paragraph{Named resolution and target exports}
For a subset of named \rir{}, we interpret instructions while retaining the nesting of a control or adjoint with one operand. We resolve register views and formal memory arguments, and check that the module graph agrees with the production resolver after semantic normalization. The soundness theorem gives equal interpretations for every snapshot and slot interpretation whenever this check succeeds. It uses the resolver's checked dependency order and well-formedness conditions.

A separate validator decodes closed OriginIR-ext exports and checks the entry's physical wire order and bank interface. For supported gate instructions without private workspace, equality of resolved instruction forms gives equality of full matrices. A second check for reversible programs at a supplied QRAM snapshot compares the action of the export and the scoped source on every public basis input with workspace initialized to zero. It proves \(U_{\mathrm{export}}J=JU_{\mathrm{source}}\), where \(J\) appends the zero workspace, and thus includes the state of the workspace at exit. Both checks are applied to actual exports. The tests change a gate angle, permute two wires, change a bank, and omit the read that clears a private flag.

\section{Expanded binding validation}
\label{sec:substitution}

In the expanded representation, an oracle environment \(\rho\) assigns a matrix \(\rho(q)\) to each resolved call \(q\). The call identifies the wire and bank arguments, so different call sites can interpret the same oracle specification on different wires and banks.

A matrix \(U\) preserves the two guard sectors when it has no entries between the basis states that enable a control and those that disable it. This condition yields \(\ctrl_P(U)\ctrl_P(V)=\ctrl_P(UV)\) by multiplication and the identity \(PU(I-P)=0\). Disjoint target wires and read-only QRAM operations that leave the guard bits fixed give instances of this condition.

Let \(\sigma(q)\) be a closed instruction sequence for call \(q\) in the entry frame. For a resolved sequence \(t\), \(\Inline_\sigma(t)\) substitutes \(\sigma(q)\) at each call, distributes control, and distributes adjoint by reversal and conjugate transpose. A linking certificate requires that every substituted sequence denote \(\rho(q)\) and that each instruction under a control preserve its guard sectors. Both interpretations use the same QRAM snapshot.

\begin{theorem}[Substitution in instruction sequences]
\label{thm:link}
If \(t\) has a linking certificate for \(\rho\), \(\sigma\), and the supplied snapshot, then both interpretations succeed and
\[
  \den{\Inline_\sigma(t)}=\den{t}_{\rho}
\]
as matrices on the entry space. Their actions agree on every input state.
\end{theorem}

\begin{proof}
Induct on the certified instructions. The call case uses the equality of the implementation and specification matrices. A primitive instruction contributes its own matrix. In the control case, guard preservation gives the equality for the product of the controlled instructions. In the adjoint case, conjugate transposition reverses the factors. The equalities compose in sequence order.
\end{proof}

The equality includes complex phase, which may be observable under control. A projected contract \(\alpha J^\dagger UJ\approx A\), with \(J\) embedding the logical space, specifies behavior in the success subspace. Reusing it under composition requires a proof about that subspace. Section~\ref{sec:lchs} gives such a quantitative example.

\subsection{Relation to decoded artifacts}
\label{sec:validator}

The verifier compares an open program and a proposed bound program. The expected bank arguments are specified independently for each affected call. This specification lists, in bound signature order, the retained bank parameters and any additional banks from the entry. Each original position occurs exactly once. In the open program, the specification changes only the bank arguments of the designated calls.

The verifier hides the selected slot definitions in the bound program and checks that its resulting caller agrees with the open program after the specified banks have been added. It then compares the complete bound sequence with the result of substituting the implementations at their resolved wire and bank arguments. The comparison uses a semantics-preserving normal form for control and adjoint. In particular, it identifies \(\ctrl_P(U)^\dagger\) with \(\ctrl_P(U^\dagger)\) and cancels repeated adjoints in pairs. It preserves angles, guard values, wire order, and bank identity.

\begin{theorem}[Binding validation for expanded programs]
\label{thm:artifact-link}
Suppose the verifier accepts an open program and a bound program under a bank-forwarding specification. Let \(t\) be the open program's resolved sequence with the specified bank arguments, and let \(\sigma\) be the implementations at their resolved call sites. If \(t\) has a linking certificate for \(\rho\), \(\sigma\), and the supplied snapshot, then the bound program denotes \(\den{t}_\rho\) on the entry space.
\end{theorem}

The proof applies Theorem~\ref{thm:link} after the two sequences are identified by the validator and the normal form. In Deutsch--Jozsa, the verifier additionally establishes the Boolean promise and readout for the supported gate and QRAM realizations.

\section{Additional replacement examples}
\subsection{Exact sparse block encoding}
\label{sec:sparse-block}

We use the adjacency matrix of a four-vertex graph with one undirected edge, between vertices $0$ and $1$. The entry has a two-bit target $x=(x_1,x_0)$ and a one-bit signal $s$. The open oracle flips $s$ by $x_1$, flagging the columns with no neighbor. The caller then flips $x_0$. A gate realization and a QRAM realization of the slot give the same public reversible action
\[
   (x_1,x_0,s)\longmapsto (x_1,1-x_0,s\oplus x_1).
\]
The verifier establishes this permutation for both bound programs on all eight basis inputs. Its correctness theorem gives equality of the full matrices. With both the input and output signal fixed to zero, the only nonzero entries are \(A_{01}=A_{10}=1\). Thus both programs encode the adjacency matrix exactly with unit normalization. Appendix~\ref{app:cases} describes the two realizations.

\subsection{Amplitude amplification}
\label{sec:amplification}

An open caller passes a target bit and a signal bit to an amplifier slot. Both implementations use a common preparation \(A\) that applies \(R_y(2)\) to the signal and leaves the target fixed. One uses a standard amplitude amplification round~\cite{brassard2000amplitude}. The other uses fixed-point amplification of degree five~\cite{yoder2014fixedpoint}. They use three and five invocations of \(A\) or its adjoint, respectively.

The contract concerns the probability of obtaining signal zero on the all-zero input. Lean proves that the preparation has success probability at most \(0.31\) and that both bound programs have success probability at least \(0.97\). The guarantee applies to this preparation and the particular generated phase sequence. The verification procedure and rejection test appear in Appendix~\ref{app:cases}.

\section{Case-study implementation and measurements}
\label{app:cases}

\paragraph{Deutsch--Jozsa}
The one-address-bit constant-zero and first-bit implementations produce readout zero and one, respectively. With XOR expanded as Hadamard--controlled-\(Z\)--Hadamard, their gate counts \((H,X,\mathrm{CZ})\) are \((3,1,0)\) and \((5,1,1)\), respectively. Neither uses a QRAM read. We count each logical QRAM read as one memory operation.

\paragraph{Control, adjoint, and bank forwarding}
A seven-wire program invokes the same lookup slot twice under a control, with the first invocation in an adjoint context. Its address views are \([3,0]\) and \([0,2]\). Through an intermediate module, each call forwards a different bank of the same shape and the same additional bank introduced by binding. The implementation uses a controlled phase, a read from the forwarded bank, and an adjointed read from the additional bank. The verifier accepts the six-instruction expansion and rejects reversing a view, selecting the other forwarded bank, omitting the forwarding specification, and dropping an original bank argument.

A second program allocates private workspace in both the intermediate module and the lookup module. It uses two address views and three banks, one introduced at binding. The verifier rejects changes to the wire order, bank identity, bank parameter order, and caller workspace. The reference execution agrees with the expected data and phase on 32 branches of a superposition of addresses and control values.

\paragraph{Timing protocol}
\label{app:timings}
We pin each verifier workload to a logical CPU of an x86-64 Intel Xeon Gold 5118 host. Each timed subprocess runs the compiled verifier on prepared files. The measured interval includes startup, parsing, verification and reporting. The runtime table, wide Deutsch--Jozsa and wide Grover use the same \texttt{--paper} batch and compiled executable. The \texttt{--scoped} batch uses that executable for both shared-graph commands. Every reported median in these batches uses three invocations after one warm-up. The bottom module contains 16 rotations, and each additional level calls its child twice. Each timed shared-graph subprocess has a 60-second wall-clock limit and a 4\,GiB virtual address-space limit. Both commands construct their usual reports, with standard output directed to the same sink. The expanded report describes the placed sequences, and the scoped report describes the module graphs. For Table~\ref{tab:lchs-wide}, each width uses a separate pinned CPU and three invocations after one warm-up. Building the supporting Lean proofs is measured separately in Section~\ref{sec:formalization-size}. The public release supports these workloads with a smaller CLI whose startup loads only the case studies included in this paper.

\paragraph{Grover}
One bound implementation applies controlled \(Z\) to the two address wires. The other reads the word at the quantum address from the table \([0,0,0,1]\), XORs it into a private flag, applies \(Z\) to the flag, and repeats the read. The entry passes the bank through the slot to the lookup module. The repeated read restores the flag to zero and leaves the required phase, including on superpositions. The diffusion circuit uses Hadamard and phase-flip gates. The verifier checks the open and bound program structures, the selected implementation, and the supplied bank contents. The same checker accepts the retained repeated caller for both providers at 2, 4, and 17 iterations. The reference simulator agrees with the probability theorem at these counts. A caller checked with a different count is rejected. The command accepts the count as its last argument. For example, the larger input is generated with \texttt{check\_grover.py --export grover --iterations 4096}. The corresponding binding command is
\begin{lstlisting}
oracq-verify check-grover qram grover/open.json grover/qram.json grover/snapshot.json 4096
\end{lstlisting} A bank image for a different address, a changed phase gate, or an omitted uncomputation fails the check.

\paragraph{Wide Grover and LCHS}
The following script generates and checks the wider programs. The two verifier examples use the output files at the indicated widths.
\begin{lstlisting}
python scripts/check_wide_cases.py --artifacts wide
oracq-verify check-grover-wide qram 16 4 wide/grover/16/open.json \
    wide/grover/16/qram.json wide/grover/16/snapshot.json
oracq-verify check-lchs-wide 16 wide/lchs/16/open.json \
    wide/lchs/16/taylor.json wide/lchs/16/product.json wide/lchs/16/mixed.json
\end{lstlisting}

\paragraph{Approximate phase provider}
The regression generates the open and bound programs for four round counts. The command below checks 64 rounds and accepts an overall error budget of \(1/8\).
\begin{lstlisting}
python scripts/check_approximate_phase.py --artifacts phase
oracq-verify check-approx-phase phase/64/open.json phase/64/bound.json \
    0.25 64 0.125
\end{lstlisting}

\paragraph{Sparse block encoding}
The gate implementation uses a controlled bit flip on the signal, and the QRAM implementation reads the bank \([0,0,1,1]\) with an XOR into that signal. For both bindings the reversible evaluator checks the induced permutation on all eight basis inputs. Its soundness theorem relates the permutation to the circuit matrix. The test with one changed bank word is rejected.

\paragraph{Amplitude amplification}
The fixed-point construction uses degree five and synthesis parameter \(0.2\). For each bound program the verifier checks that the supplied preparation is reachable and has the same resolved instructions in the bound program. Rational approximation bounds two entries of the original preparation and a success amplitude of each amplifier. Lean transfers those bounds to the corresponding complex matrices. The test that binds the slot to the preparation alone is rejected by the success-probability check.

\section{Numerical certification of the LCHS circuits}
\label{app:lchs}

\paragraph{Common interface and checked assembly}
Each slot has one target and seven signal bits. The adapter calls a branch on the prefix it uses and applies \(R_y(2\arccos(\alpha_j/2))\) to the final signal bit, where \(\alpha_j\) is the implementation's original normalization. The ideal zero-to-zero amplitude of the extra flag is \(\alpha_j/2\). The numerical certificate checks the actual decimal angle in the bound artifact. The outer normalization is the exact decimal \(0.9867606471697511\). Verification compares the open and bound module graphs, transfers the selector gate evidence when the resolved word agrees, and checks the scale and success block of each selected branch. The exact word comparison and the ensuing coefficient bound establish the assembly discrepancy as described in Section~\ref{sec:lchs}.

\paragraph{Reproduction}
The following commands export and check one open program and three candidate bindings.
\begin{lstlisting}
python scripts/check_replacements.py --export lchs
oracq-verify check-lchs-link lchs/lchs-open.json \
  lchs/lchs-bound-taylor.json lchs/lchs-bound-product.json \
  lchs/lchs-bound-mixed.json
\end{lstlisting}
The Lean theorem \texttt{LCHS.LinkedAssembly.sound} relates an accepted binding to its decoded circuit and its finite-quadrature block-encoding guarantee. The SELECT/PREPARE denotation follows from \texttt{SelectLayout.word\_matrix}. The proof relates the resulting dependent-width matrix to the physical success frame on the ten decoded wires.

\paragraph{Standalone assemblies}
The standalone Taylor and product-formula circuits, in which normalization and signal width are chosen independently, use nine and three wires respectively. Both scaled blocks approximate \(e^{-t(L+iH)}\) to within \(0.47\) in operator norm and differ by at most \(10^{-3}\). For the Taylor circuit, the scaled block differs by at most \(10^{-11}\) from \(\sum_jc_j(I-it(H+k_jL)-t^2(H+k_jL)^2/2)\). Its success probability on the prepared input is at least \(36/49\).

\paragraph{Circuits and normalizations}
The standalone Taylor circuit has 32 modules and expands to 155 instructions, of which 152 are controlled and 33 are adjointed. Both circuits use wire zero as the target. The Taylor and product-formula normalizations are the exact decimals \(0.551563578415394\) and \(0.49338032358487555\), respectively. The verifier follows the four controlled calls in the serialized coefficient selection and decodes their branches individually. The Taylor branch at \(k=-1\) has seven wires and normalization \(2545/2048\). The product formula uses Hadamard gates and conditional and unconditional phase operations. Oracq selects the branch implementation during construction through its \texttt{hamiltonian\_function} argument.

\paragraph{Rational evaluation}
For each of the two logical input basis states, the verifier evaluates the corresponding circuit column and extracts the two entries with signal zero. It approximates each phase by a rational exponential partial sum. It rounds the state vector after each gate and uses unitarity to bound the total error by summing the per-gate errors. For the comparison with \(T\), multiplication by \(\pi\) makes the four reference entries Gaussian rationals. A rational interval for \(\pi\) then bounds the error introduced in comparing with the scaled success block. Two closed rational comparisons for the full Taylor circuit are discharged using Lean's compiled decision procedure. The error propagation and the identification of the exact matrix are proved in Lean. The separate seven-wire branch uses direct kernel evaluation for its closed comparisons.

\paragraph{Continuum and pairwise bounds}
Each quadrature node satisfies \(\|t(H+k_jL)\|\le 7/32\), with degree-two truncation bound \(343/98304\). The weighted bound for this error is \((\sum_j |c_j|)\,343/98304\). Expanding the continuum operator to degree two gives the same truncation bound. In the difference of its Taylor polynomial and \(T\), the diagonal entries have moduli at most \(0.452\) and the off-diagonal entries at most \(0.013\). Writing the latter as a diagonal matrix times a bit flip gives an operator-norm bound of \(0.465\) on the polynomial difference. This yields the stated bound of \(0.47\) for the scaled Taylor circuit and \(0.48\) for the exact four-node quadrature. In the comparison of the two implementations, the verifier compares their scaled blocks with the continuum polynomial and also bounds their difference directly. In each comparison the entry bounds include the phase approximation and rounding errors relative to the exact serialized angles.

\end{document}